# Merging-Diverging Hybrid Transformer Networks for Survival Prediction in Head and Neck Cancer

Mingyuan Meng[1,2], Lei Bi[2], Michael Fulham[1,3], Dagan Feng[1,4], and Jinman Kim[1]

[1] School of Computer Science, The University of Sydney, Sydney, Australia.
[2] Institute of Translational Medicine, Shanghai Jiao Tong University, Shanghai, China.
[3] Department of Molecular Imaging, Royal Prince Alfred Hospital, Sydney, Australia.
[4] Med-X Research Institute, Shanghai Jiao Tong University, Shanghai, China.
lei.bi@sjtu.edu.cn

**Abstract.** Survival prediction is crucial for cancer patients as it provides early prognostic information for treatment planning. Recently, deep survival models based on deep learning and medical images have shown promising performance for survival prediction. However, existing deep survival models are not well developed in utilizing multi-modality images (e.g., PET-CT) and in extracting region-specific information (e.g., the prognostic information in Primary Tumor (PT) and Metastatic Lymph Node (MLN) regions). In view of this, we propose a merging-diverging learning framework for survival prediction from multi-modality images. This framework has a merging encoder to fuse multi-modality information and a diverging decoder to extract region-specific information. In the merging encoder, we propose a Hybrid Parallel Cross-Attention (HPCA) block to effectively fuse multi-modality features via parallel convolutional layers and cross-attention transformers. In the diverging decoder, we propose a Region-specific Attention Gate (RAG) block to screen out the features related to lesion regions. Our framework is demonstrated on survival prediction from PET-CT images in Head and Neck (H&N) cancer, by designing an X-shape merging-diverging hybrid transformer network (named XSurv). Our XSurv combines the complementary information in PET and CT images and extracts the region-specific prognostic information in PT and MLN regions. Extensive experiments on the public dataset of HEad and neCK TumOR segmentation and outcome prediction challenge (HECKTOR 2022) demonstrate that our XSurv outperforms state-of-the-art survival prediction methods.

## 1 Introduction

Head and Neck (H&N) cancer refers to malignant tumors in H&N regions, which is among the most common cancers worldwide [1]. Survival prediction, a regression task that models the survival outcomes of patients, is crucial for H&N cancer patients: it provides early prognostic information to guide treatment planning and potentially improves the overall survival outcomes of patients [2]. Multi-modality imaging of Positron Emission Tomography – Computed Tomography (PET-CT) has been shown to benefit survival prediction as it offers both anatomical (CT) and metabolic (PET) information about tumors [3, 4]. Therefore, survival prediction from PET-CT images in H&N cancer has attracted wide attention and serves as a key research area. For instance, HEad and neCK TumOR segmentation and outcome prediction challenges (HECKTOR) have been held for the last three years to facilitate the development of new algorithms for survival prediction from PET-CT images in H&N cancer [5-7].

Traditional survival prediction methods are usually based on radiomics [8], where handcrafted radiomics features are extracted from pre-segmented tumor regions and then are modeled by statistical survival models, such as the Cox Proportional Hazard (CoxPH) model [9]. In addition, deep survival models based on deep learning have been proposed to perform end-to-end survival prediction from medical images, where pre-segmented tumor masks are often unrequired [10]. Deep survival models usually adopt Convolutional Neural Networks (CNNs) to extract image features, and recently Visual Transformers (ViT) have been adopted for its capabilities to capture long-range dependency within images [11, 12]. These deep survival models have shown the potential to outperform traditional survival prediction methods [13]. For survival prediction in H&N cancer, deep survival models have achieved top performance in the HECKTOR 2021/2022 and are regarded as state-of-the-art [14-16]. Nevertheless, we identified that existing deep survival models still have two main limitations.

Firstly, existing deep survival models are underdeveloped in utilizing complementary multi-modality information, such as the metabolic and anatomical information in PET and CT images. For survival prediction in H&N cancer, existing methods usually use single imaging modality [17, 18] or rely on early fusion (i.e., concatenating multi-modality images as multi-channel inputs) to combine multi-modality information [11, 14-16, 19]. In addition, late fusion has been used for survival prediction in other diseases such as gliomas and tuberculosis [20, 21], where multi-modality features were extracted by multiple independent encoders with resultant features fused. However, early fusion has difficulties in extracting intra-modality information due to entangled (concatenated) images for feature extraction, while late fusion has difficulties in extracting inter-modality information due to fully independent feature extraction. Recently, Tang et al. [22] attempted to address this limitation by proposing a Multi-scale Non-local Attention Fusion (MNAF) block for survival prediction of glioma patients, in which multi-modality features were fused via non-local attention mechanism [23] at multiple scales. However, the performance of this method heavily relies on using tumor segmentation masks as inputs, which limits its generalizability.

Secondly, although deep survival models have advantages in performing end-to-end survival prediction without requiring tumor masks, this also incurs difficulties in extracting region-specific information, such as the prognostic information in Primary Tumor (PT) and Metastatic Lymph Node (MLN) regions. To address this limitation, recent deep survival models adopted multi-task learning for joint tumor segmentation and survival prediction, to implicitly guide the model to extract features related to tumor regions [11, 16, 24-26]. However, most of them only considered PT segmentation and ignored the prognostic information in MLN regions [11, 24-26]. Meng et al. [16] performed survival prediction with joint PT-MLN segmentation and achieved one of the top performances in HECKTOR 2022. However, this method extracted entangled features related to both PT and MLN regions, which incurs difficulties in discovering the prognostic information in PT-/MLN-only regions.

In this study, we design an X-shape merging-diverging hybrid transformer network (named XSurv, Fig. 1) for survival prediction in H&N cancer. Our XSurv has a merging encoder to fuse complementary anatomical and metabolic information in PET and CT images and has a diverging decoder to extract region-specific prognostic information in PT and MLN regions. Our technical contributions in XSurv are three folds: (i) We propose a merging-diverging learning framework for survival prediction. This framework is specialized in leveraging multi-modality images and extracting region-specific information, which potentially could be applied to many survival prediction tasks with multi-modality imaging. (ii) We propose a Hybrid Parallel Cross-Attention (HPCA) block for multi-modality feature learning, where both local intra-modality and global inter-modality features are learned via parallel convolutional layers and cross-attention transformers. (iii) We propose a Region-specific Attention Gate (RAG) block for region-specific feature extraction, which screens out the features related to lesion regions. Extensive experiments on the public dataset of HECKTOR 2022 [7] demonstrate that our XSurv outperforms state-of-the-art survival prediction methods, including the top-performing methods in HECKTOR 2022.

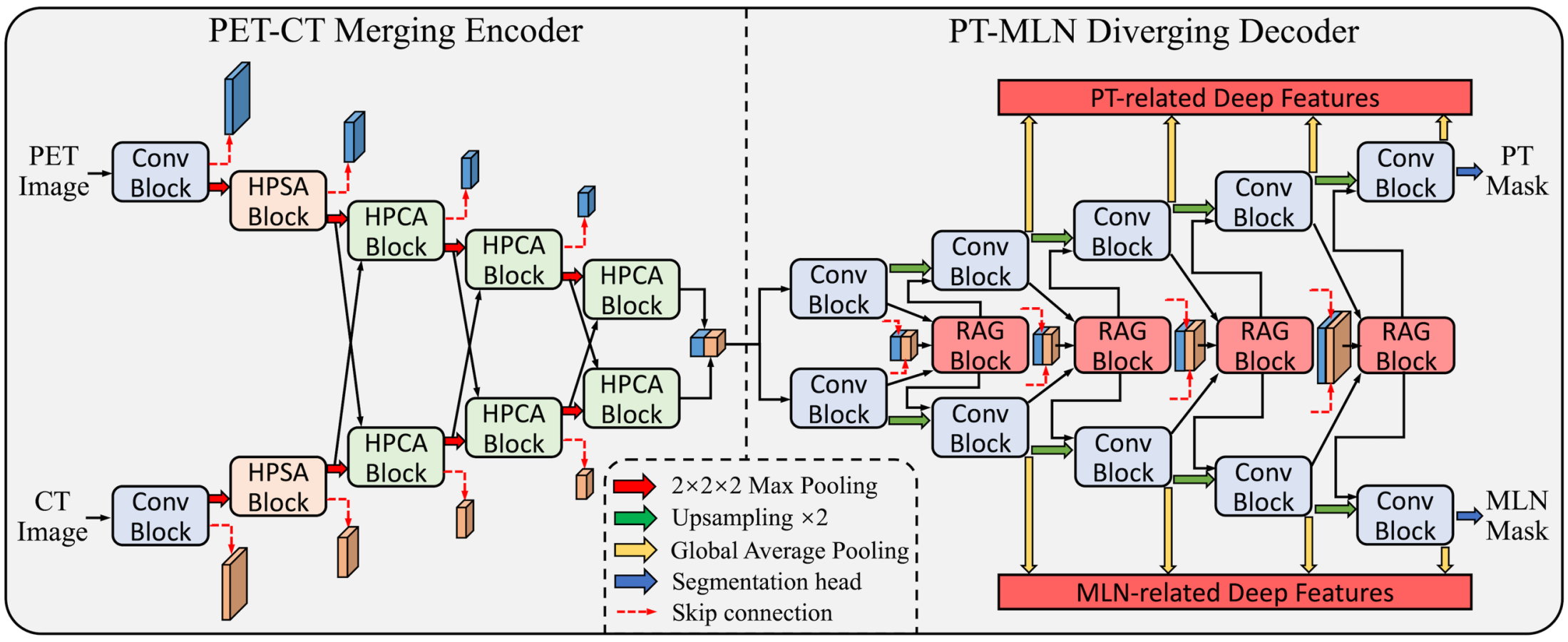

**Fig. 1.** The architecture of our XSurv. The architecture parameters $N_{conv}$, $N_{self}$, and $N_{cross}$ are set as 1, 1, and 3 for illustration. Survival prediction head is omitted here for clarity.

## 2 Method

Fig. 1 illustrates the overall architecture of our XSurv, which presents an X-shape architecture consisting of a merging encoder for multi-modality feature learning and a diverging decoder for region-specific feature extraction. The encoder includes two PET-/CT-specific feature learning branches with HPCA blocks (refer to Section 2.1), while the decoder includes two PT-/MLN-specific feature extraction branches with RAG blocks (refer to Section 2.2). Our XSurv performs joint survival prediction and segmentation, where the two decoder branches are trained to perform PT/MLN seg-mentation and provide PT-/MLN-related deep features for survival prediction (refer to Section 2.3). Our XSurv also can be enhanced by leveraging the radiomics features extracted from the XSurv-segmented PT/MLN regions (refer to Section 2.4). Our implementation is provided at https://github.com/MungoMeng/Survival-XSurv.

## 2.1 PET-CT Merging Encoder

Assuming $N_{conv}$, $N_{self}$, and $N_{cross}$ are three architecture parameters, each encoder branch consists of $N_{conv}$ Conv blocks, $N_{self}$ Hybrid Parallel Self-Attention (HPSA) blocks, and $N_{cross}$ HPCA blocks. Max pooling is applied between blocks and the features before max pooling are propagated to the decoder through skip connections. As shown in Fig. 2(a), HPCA blocks perform parallel convolution and cross-attention operations. The convolution operations are realized using successive convolutional layers with residual connections, while the cross-attention operations are realized using Swin Transformer [27] where the input $x_{in}$ (from the same encoder branch) is projected as $Q$ and the input $x_{cross}$ (from the other encoder branch) is projected as $K$ and $V$. In addition, Conv blocks perform the same convolution operations as HPCA blocks but discard cross-attention operations; HPSA blocks share the same overall architecture with HPCA blocks but perform self-attention within the input $x_{in}$ (i.e., the $x_{in}$ is projected as $Q$, $K$ and $V$). Conv and HPSA blocks are used first and then followed by HPCA blocks, which enables the XSurv to learn both intra- and inter-modality information. In this study, we set $N_{conv}$, $N_{self}$, and $N_{cross}$ as 1, 1, and 3, as this setting achieved the best validation results (refer to the supplementary materials). Other architecture details are also presented in the supplementary materials.

The idea of adopting convolutions and transformers in parallel has been explored for segmentation [28], which suggests that parallelly aggregating global and local information is beneficial for feature learning. In this study, we extend this idea to multi-modality feature learning, which parallelly aggregates global inter-modality and local intra-modality information via HPCA blocks, to discover inter-modality interactions while preserving intra-modality characteristics.

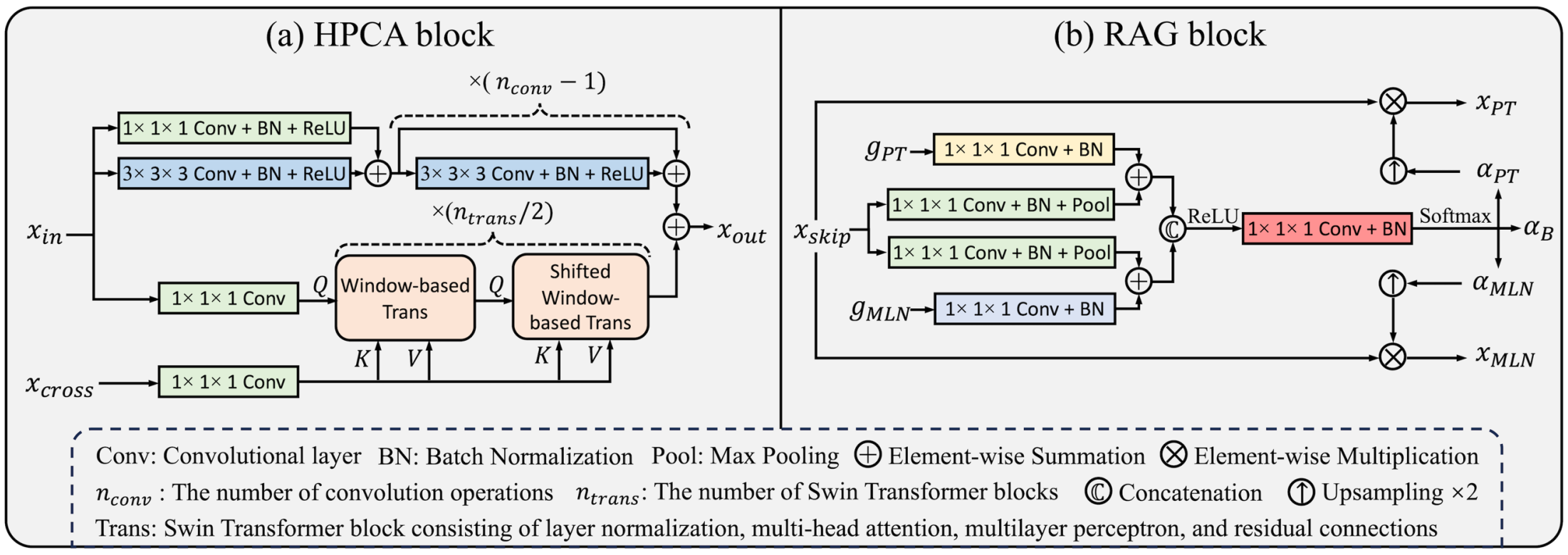

**Fig. 2.** The detailed architecture of the proposed (a) Hybrid Parallel Cross-Attention (HPCA) block and (b) Region-specific Attention Gate (RAG) block.

## 2.2 PT-MLN Diverging Decoder

As shown in Fig. 1, each decoder branch is symmetric to the encoder branch and thus includes a total of ($N_{conv}$+$N_{self}$+$N_{cross}$) Conv blocks. The features propagated from skip connections are fed into RAG blocks for feature diverging before entering the Conv blocks in two decoder branches, where the output of the former Conv block is upsampled and concatenated with the output of the RAG block. As shown in Fig. 2(b), RAG blocks generate three softmax-activated spatial attention maps $\alpha_{PT}$, $\alpha_{MLN}$, and $\alpha_B$ that correspond to PT, MLN, and background regions. These attention maps are computed based on the contextual information provided by the gating signals $g_{PT}$ and $g_{MLN}$ (which are the outputs of the former Conv blocks in the PT and MLN branches). The attention maps $\alpha_{PT}$ and $\alpha_{MLN}$ are multiplied with the features $x_{skip}$ that are propagated from skip connections, which spatially diverge the features $x_{skip}$ into PT- and MLN-related features $x_{PT}$ and $x_{MLN}$. Different from the vanilla Attention Gate (AG) block [29], RAG blocks leverage the gating signals from two decoder branches and generate mutually exclusive (softmax-activated) attention maps.

The output of the last Conv block in the PT/MLN branch is fed into a segmentation head, which generates PT/MLN segmentation masks using a sigmoid-activated 1×1×1 convolutional layer. In addition, the outputs of all but not the first Conv blocks in the PT/MLN branches are fed into global averaging pooling layers to derive PT-/MLN-related deep features. Finally, all deep features are fed into a survival prediction head, which maps the deep features into a survival score using two fully-connected layers with dropout, L2 regularization, and sigmoid activation.

## 2.3 Multi-task Learning

Following existing multi-task deep survival models [11, 16, 24-26], our XSurv is end-to-end trained for survival prediction and PT-MLN segmentation using a combined loss: $\mathcal{L} = \mathcal{L}_{Surv} + \lambda(\mathcal{L}_{PT} + \mathcal{L}_{MLN})$, where the $\lambda$ is a parameter to balance the survival prediction term $\mathcal{L}_{Surv}$ and the PT/MLN segmentation terms $\mathcal{L}_{PT/MLN}$. We follow [15] to adopt a negative log-likelihood loss [30] as the $\mathcal{L}_{Surv}$. For the $\mathcal{L}_{PT/MLN}$, we adopt the sum of Dice [31] and Focal [32] losses. The loss functions are detailed in the supplementary materials. The $\lambda$ is set as 1 in the experiments as default.

## 2.4 Radiomics Enhancement

Our XSurv also can be enhanced by leveraging radiomics features (denoted as Radio-XSurv). Following [16], radiomics features are extracted from the XSurv-segmented PT/MLN regions via Pyradiomics [33] and selected by Least Absolute Shrinkage and Selection Operator (LASSO) regression. The process of radiomics feature extraction is provided in the supplementary materials. Then, a CoxPH model [9] is adopted to integrate the selected radiomics features and the XSurv-predicted survival score to make the final prediction. In addition, clinical indicators (e.g., age, gender) also can be integrated by the CoxPH model.

# 3 Experimental Setup

## 3.1 Dataset and Preprocessing

We adopted the training dataset of HECKTOR 2022 (refer to https://hecktor.grand-challenge.org/), including 488 H&N cancer patients acquired from seven medical centers [7], while the testing dataset was excluded as its ground-truth labels are not released. Each patient underwent pretreatment PET/CT and has clinical indicators. We present the distributions of all clinical indicators in the supplementary materials. Recurrence-Free Survival (RFS), including time-to-event in days and censored-or-not status, was provided as ground truth for survival prediction, while PT and MLN annotations were provided for segmentation. The patients from two centers (CHUM and CHUV) were used for testing and other patients for training, which split the data into 386/102 patients in training/testing sets. We trained and validated models using 5-fold cross-validation within the training set and evaluated them in the testing set.

We resampled PET-CT images into isotropic voxels where 1 voxel corresponds to 1 $mm^3$. Each image was cropped to 160×160×160 voxels with the tumor located in the center. PET images were standardized using Z-score normalization, while CT images were clipped to [−1024, 1024] and then mapped to [−1, 1]. In addition, we performed univariate and multivariate Cox analyses on the clinical indicators to screen out the prognostic indicators with significant relevance to RFS ($P$<0.05).

## 3.2 Implementation Details

We implemented our XSurv using PyTorch on a 12GB GeForce GTX Titan X GPU. Our XSurv was trained for 12,000 iterations using an Adam optimizer with a batch size of 2. Each training batch included the same number of censored and uncensored samples. The learning rate was set as 1e-4 initially and then reset to 5e−5 and 1e−5 at the 4,000$^{th}$ and 8,000$^{th}$ training iteration. Data augmentation was applied in real-time during training to minimize overfitting, including random affine transformations and random cropping to 112×112×112 voxels. Validation was performed after every 200 training iterations and the model achieving the highest validation result was preserved. In our experiments, one training iteration (including data augmentation) took roughly 4.2 second, and one inference iteration took roughly 0.61 second.

## 3.3 Experimental Settings

We compared our XSurv to six state-of-the-art survival prediction methods, including two traditional radiomics-based methods and four deep survival models. The included traditional methods are CoxPH [9] and Individual Coefficient Approximation for Risk Estimation (ICARE) [34]. For traditional methods, radiomics features were extracted from the provided ground-truth tumor regions and selected by LASSO regression. The included deep survival models are Deep Multi-Task Logistic Regression and CoxPH ensemble (DeepMTLR-CoxPH) [14], Transformer-based Multimodal net-works for Segmentation and Survival prediction (TMSS) [11], Deep Multi-task Survival model (DeepMTS) [24], and Radiomics-enhanced DeepMTS (Radio-DeepMTS) [16]. DeepMTLR-CoxPH, ICARE, and Radio-DeepMTS achieved top performance in HECKTOR 2021 and 2022. For a fair comparison, all methods took the same preprocessed images and clinical indicators as inputs. Survival prediction and segmentation were evaluated using Concordance index (C-index) and Dice Similarity Coefficient (DSC), which are the standard evaluation metrics in the challenges [6, 7, 35].

We also performed two ablation studies on the encoder and decoder separately: (i) We replaced HPCA/HPSA blocks with Conv blocks and compared different strategies to combine PET-CT images. (ii) We removed RAG blocks and compared different strategies to extract PT/MLN-related information.

**Table 1.** Comparison between XSurv and state-of-the-art survival prediction methods.

| Methods | | Survival prediction (C-index) | PT segmentation (DSC) | MLN segmentation (DSC) |
|---|---|---|---|---|
| CoxPH [9] | Radiomics | 0.745±0.024* | / | / |
| ICARE [34] | Radiomics | 0.765±0.019* | / | / |
| DeepMTLR-CoxPH [14] | CNN | 0.748±0.025* | / | / |
| TMSS [11] | ViT+CNN | 0.761±0.028* | 0.784±0.015* | 0.724±0.018* |
| DeepMTS [24] | CNN | 0.757±0.022* | 0.754±0.010* | 0.715±0.013* |
| XSurv (Ours) | Hybrid | 0.782±0.018 | **0.800±0.006** | **0.754±0.008** |
| Radio-DeepMTS [16] | CNN+Radiomics | 0.776±0.018‡ | 0.754±0.010‡ | 0.715±0.013‡ |
| Radio-XSurv (Ours) | Hybrid+Radiomics | **0.798±0.015** | **0.800±0.006** | **0.754±0.008** |

**Bold**: the best result in each column is in bold. ±: standard deviation.
*: $P$<0.05, in comparison to XSurv. ‡: $P$<0.05, in comparison to Radio-XSurv.

# 4 Results and Discussion

The comparison between our XSurv and the state-of-the-art methods is presented in Table 1. Our XSurv achieved a higher C-index than all compared methods, which demonstrates that our XSurv has achieved state-of-the-art performance in survival prediction of H&N cancer. When radiomics enhancement was adopted in XSurv and DeepMTS, our Radio-XSurv also outperformed the Radio-DeepMTS and achieved the highest C-index. Moreover, the segmentation results of multi-task deep survival models (TMSS, DeepMTS, and XSurv) are also reported in Table 1. Our XSurv achieved higher DSCs than TMSS and DeepMTS, which demonstrates that our XSurv can locate PT and MLN more precisely and this infers that our XSurv has better learning capability. We attribute these performance improvements to the use of our proposed merging-diverging learning framework, HPCA block, and RAG block, which can be evidenced by ablation studies.

The ablation study on the PET-CT merging encoder is shown in Table 2. We found that using PET alone resulted in a higher C-index than using both PET-CT with early or late fusion. This finding is consistent with Wang et al. [19]'s study, which suggests that early and late fusion cannot effectively leverage the complementary information in PET-CT images. As we have mentioned, early and late fusion have difficulties in extracting intra- and inter-modality information, respectively. Our encoder first adopts

Conv/HPSA blocks to extract intra-modality information and then leverages HPCA blocks to discover their interactions, which achieved the highest C-index. For PT and MLN segmentation, our encoder also achieved the highest DSCs, which indicates that our encoder also can improve segmentation. In addition, MNAF blocks [22] were compared and showed poor performance. This is likely attributed to the fact that leveraging non-local attention at multiple scales has corrupted local spatial information, which degraded the segmentation performance and distracted the model from PT and MLN regions. To relieve this problem, in Tang et al.'s study [22], tumor segmentation masks were fed into the model as explicit guidance to tumor regions. However, it is intractable to have segmentation masks at the inference stage in clinical practice.

The ablation study on the PT-MLN diverging decoder is shown in Table 3. We found that, even without adopting AG, using a dual-branch decoder for PT and MLN segmentation resulted in a higher C-index than using a single-branch decoder, which demonstrates the effectiveness of our diverging decoder design. Adopting vanilla AG [29] or RAG in the dual-branch decoder further improved survival prediction. Compared to the vanilla AG, our RAG contributed to a larger improvement, and this enabled our decoder to achieve the highest C-index. In the supplementary materials, we visualized the attention maps produced by RAG blocks, where the attention maps can precisely locate PT/MLN regions and screen out PT-/MLN-related features. For PT and MLN segmentation, using a single-branch decoder for PT- or MLN-only segmentation achieved the highest DSCs. This is expected as the model can leverage all its capabilities to segment only one target. Nevertheless, our decoder still achieved the second-best DSCs in both PT and MLN segmentation with a small gap.

**Table 2.** Ablation study on the PET-CT merging encoder.

<table>
<tr><th colspan="2">Methods</th><th>Survival prediction (C-index)</th><th>PT segmentation (DSC)</th><th>MLN segmentation (DSC)</th></tr>
<tr><td rowspan="3">SBE with $C_e$=[16, 32, 64, 128, 256]</td><td>Only PET</td><td>0.767</td><td>0.753</td><td>0.699</td></tr>
<tr><td>Only CT</td><td>0.637</td><td>0.630</td><td>0.702</td></tr>
<tr><td>Early fusion</td><td>0.755</td><td>0.783</td><td>0.722</td></tr>
<tr><td rowspan="3">DBE with $C_e$=[8, 16, 32, 64, 128]</td><td>Late fusion</td><td>0.762</td><td>0.796</td><td>0.744</td></tr>
<tr><td>MNAF [22]</td><td>0.688</td><td>0.741</td><td>0.683</td></tr>
<tr><td>Ours</td><td>0.782</td><td>0.800</td><td>0.754</td></tr>
</table>

**Bold**: the best result in each column is in bold. SBE: single-branch encoder. DBE: dual-branch encoder. $C_e$: the channel numbers or embedding dimensions used in the encoder.

**Table 3.** Ablation study on the PT-MLN diverging decoder.

<table>
<tr><th colspan="2">Methods</th><th>Survival prediction (C-index)</th><th>PT segmentation (DSC)</th><th>MLN segmentation (DSC)</th></tr>
<tr><td rowspan="3">SBD with $C_d$=[256, 128, 64, 32, 16]</td><td>Only PT</td><td>0.751</td><td>0.803</td><td>/</td></tr>
<tr><td>Only MLN</td><td>0.746</td><td>/</td><td>0.758</td></tr>
<tr><td>PT and MLN</td><td>0.765</td><td>0.790</td><td>0.734</td></tr>
<tr><td rowspan="3">DBD with $C_d$=[128, 64, 32, 16, 8]</td><td>No AG</td><td>0.770</td><td>0.792</td><td>0.740</td></tr>
<tr><td>Vanilla AG [29]</td><td>0.774</td><td>0.795</td><td>0.745</td></tr>
<tr><td>Ours</td><td>0.782</td><td>0.800</td><td>0.754</td></tr>
</table>

**Bold**: the best result in each column is in bold. SBD: single-branch decoder. DBD: dual-branch decoder. $C_d$: the channel numbers used in the decoder.

## 5 Conclusion

We have outlined an X-shape merging-diverging hybrid transformer network (XSurv) for survival prediction from PET-CT images in H&N cancer. Within the XSurv, we propose a merging-diverging learning framework, a Hybrid Parallel Cross-Attention (HPCA) block, and a Region-specific Attention Gate (RAG) block, to learn complementary information from multi-modality images and extract region-specific prognostic information for survival prediction. Extensive experiments have shown that the proposed framework and blocks enable our XSurv to outperform state-of-the-art survival prediction methods on the well-benchmarked HECKTOR 2022 dataset.

**Acknowledgement.** This work was supported by Australian Research Council (ARC) under Grant DP200103748.

## References

1. Parkin, D.M., Bray, F., Ferlay, J., Pisani, P.: Global cancer statistics, 2002. CA: a cancer journal for clinicians 55(2), 74–108 (2005).
2. Wang, X., Li, B.B.: Deep learning in head and neck tumor multiomics diagnosis and analysis: review of the literature. Frontiers in Genetics 12, 624820 (2021).
3. Bogowicz, M., et al.: Comparison of PET and CT radiomics for prediction of local tumor control in head and neck squamous cell carcinoma. Acta oncologica 56(11), 1531-1536 (2017).
4. Gu, B., et al.: Prediction of 5-year progression-free survival in advanced nasopharyngeal carcinoma with pretreatment PET/CT using multi-modality deep learning-based radiomics. Frontiers in oncology 12, 899351 (2022).
5. Andrearczyk, V. et al.: Overview of the HECKTOR Challenge at MICCAI 2020: Automatic Head and Neck Tumor Segmentation in PET/CT. In: Andrearczyk, V., et al. (eds.) HECKTOR 2020. LNCS, vol. 12603, pp. 1-21. Springer, Cham (2021).
6. Andrearczyk, V. et al.: Overview of the HECKTOR Challenge at MICCAI 2021: Automatic Head and Neck Tumor Segmentation and Outcome Prediction in PET/CT Images. In: Andrearczyk, V., et al. (eds.) HECKTOR 2021. LNCS, vol. 13209, pp. 1-37. Springer, Cham (2022).
7. Andrearczyk, V., et al. Overview of the HECKTOR Challenge at MICCAI 2022: Automatic Head and Neck Tumor Segmentation and Outcome Prediction in PET/CT. In: Andrearczyk, V., et al. (eds.) HECKTOR 2022. LNCS, vol. 13626, pp. 1-30. Springer, Cham (2023).
8. Gillies, R. J., Kinahan, P. E., Hricak, H.: Radiomics: Images are more than pictures, they are data. Radiology 278(2), 563–577 (2016).
9. Cox, D. R.: Regression models and life-tables. Journal of the Royal Statistical Society: Series B (Methodological) 34(2), 187-202 (1972).
10. Deepa, P., Gunavathi, C.: A systematic review on machine learning and deep learning techniques in cancer survival prediction. Progress in Biophysics and Molecular Biology 174, 62-71 (2022).
11. Saeed, N., Sobirov, I., Al Majzoub, R., Yaqub, M.: TMSS: An End-to-End Transformer-Based Multimodal Network for Segmentation and Survival Prediction. In: Wang, L., et al. (eds.) MICCAI 2022. LNCS, vol. 13437, pp. 319-329. Springer, Cham (2022).

12. Zheng, H., et al.: Multi-transSP: Multimodal Transformer for Survival Prediction of Nasopharyngeal Carcinoma Patients. In: Wang, L., et al. (eds.) MICCAI 2022. LNCS, vol. 13437, pp. 234-243. Springer, Cham (2022).
13. Afshar, P., et al.: From handcrafted to deep-learning-based cancer radiomics: challenges and opportunities. IEEE Signal Processing Magazine 36(4), 132-160 (2019).
14. Saeed, N., et al. An ensemble approach for patient prognosis of head and neck tumor using multimodal data. In: Andrearczyk, V., et al. (eds.) HECKTOR 2021. LNCS, vol. 13209, pp. 278-286. Springer, Cham (2022).
15. Naser, M.A., et al.: Progression free survival prediction for head and neck cancer using deep learning based on clinical and PET/CT imaging data. In: Andrearczyk, V., et al. (eds.) HECKTOR 2021. LNCS, vol. 13209, pp. 287-299. Springer, Cham (2022).
16. Meng, M., Bi, L., Feng, D., Kim, J.: Radiomics-enhanced Deep Multi-task Learning for Outcome Prediction in Head and Neck Cancer. In: Andrearczyk, V., et al. (eds.) HECKTOR 2022. LNCS, vol. 13626, pp. 135-143. Springer, Cham (2023).
17. Diamant, A., et al.: Deep learning in head & neck cancer outcome prediction. Scientific reports 9, 2764 (2019).
18. Fujima, N., et al.: Prediction of the local treatment outcome in patients with oropharyngeal squamous cell carcinoma using deep learning analysis of pretreatment FDG-PET images. BMC Cancer 21, 900 (2021).
19. Wang, Y., et al.: Deep learning based time-to-event analysis with PET, CT and joint PET/CT for head and neck cancer prognosis. Computer Methods and Programs in Biomedicine 222, 106948 (2022).
20. Zhou, T., et al.: M^2Net: Multi-modal Multi-channel Network for Overall Survival Time Prediction of Brain Tumor Patients. In: Martel, A.L., et al. (eds.) MICCAI 2020. LNCS, vol. 12263, pp. 221–231. Springer, Cham (2020).
21. D'Souza, N.S., et al.: Fusing Modalities by Multiplexed Graph Neural Networks for Outcome Prediction in Tuberculosis. In: Wang, L., et al. (eds.) MICCAI 2022. LNCS, vol. 13437, pp. 287-297. Springer, Cham (2022).
22. Tang, W., et al.: MMMNA-Net for Overall Survival Time Prediction of Brain Tumor Patients. In: Annual International Conference of the IEEE Engineering in Medicine & Biology Society, pp. 3805-3808 (2022).
23. Wang, X., Girshick, R., Gupta, A., He, K.: Non-local neural networks. In: IEEE conference on Computer Vision and Pattern Recognition, pp. 7794-7803 (2018).
24. Meng, M., et al.: DeepMTS: Deep multi-task learning for survival prediction in patients with advanced nasopharyngeal carcinoma using pretreatment PET/CT. IEEE Journal of Biomedical and Health Informatics 26(9), 4497-4507 (2022).
25. Meng, M., Peng, Y., Bi, L., Kim, J.: Multi-task deep learning for joint tumor segmentation and outcome prediction in head and neck cancer. In: Andrearczyk, V., et al. (eds.) HECKTOR 2021. LNCS, vol. 13209, pp. 160-167. Springer, Cham (2022).
26. Andrearczyk, V., et al.: Multi-task deep segmentation and radiomics for automatic prognosis in head and neck cancer. In: Rekik, I., et al. (eds.) PRIME 2021. LNCS, vol. 12928, pp. 147-156. Springer, Cham (2022).
27. Liu, Z., et al.: Swin transformer: Hierarchical vision transformer using shifted windows. In: IEEE/CVF International Conference on Computer Vision, pp. 10012-10022 (2021).
28. Liu, W., et al.: PHTrans: Parallelly aggregating global and local representations for medical image segmentation. In: Wang, L., et al. (eds.) MICCAI 2022. LNCS, vol. 13435, pp. 235-244. Springer, Cham (2022).
29. Schlemper, J., et al.: Attention gated networks: Learning to leverage salient regions in medical images. Medical image analysis 53, 197-207 (2019).

30. Gensheimer, M.F., Narasimhan, B.: A scalable discrete-time survival model for neural networks. PeerJ 7, e6257 (2019).
31. Milletari, F., et al.: V-Net: fully convolutional neural networks for volumetric medical image segmentation. In: International Conference on 3D Vision, pp. 565-571 (2016).
32. Lin, T.Y., et al.: Focal loss for dense object detection. In: IEEE International Conference on Computer Vision, pp. 2980-2988 (2017).
33. Van Griethuysen, J.J., et al.: Computational radiomics system to decode the radiographic phenotype. Cancer research 77(21), e104-e107 (2017).
34. Rebaud, L., et al.: Simplicity is All You Need: Out-of-the-Box nnUNet followed by Binary-Weighted Radiomic Model for Segmentation and Outcome Prediction in Head and Neck PET/CT. In: Andrearczyk, V., et al. (eds.) HECKTOR 2022. LNCS, vol. 13626, pp. 121-134. Springer, Cham (2023).
35. Eisenmann, M., et al.: Biomedical image analysis competitions: The state of current participation practice. arXiv preprint, arXiv:2212.08568 (2022).

# Supplementary Materials for "Merging-Diverging Hybrid Transformer Networks for Survival Prediction in Head and Neck Cancer"

## A. Loss Function Details

The survival prediction loss term $\mathcal{L}_{Surv}$ is a negative log-likelihood function:

$$\mathcal{L}_{Surv} = -\sum_{i=1}^{N} \log\left(\max\left(1 + S_i\left(S_i^{pred} - 1\right), \varepsilon\right)\right) + \log\left(\max\left(1 - \bar{S}_i S_i^{pred}, \varepsilon\right)\right), \quad (1)$$

where $S^{pred}, S, \bar{S} \in \mathbb{R}^N$. The $S^{pred}$ is the output of the survival prediction head, which represent the conditional probabilities of patient surviving in $N$ time intervals. In this study, we set $N = 10$, where the RFS of all training samples are evenly distributed in 10 intervals. $S$ and $\bar{S}$ are two label vectors generated from RFS labels (time-to-event in days and censored-or-not status): For $S$, all the time intervals preceding the events are set to 1 while other are 0; For $\bar{S}$, the time interval with event occurring (for uncensored patients only) is set to 1 while other are 0. With the predicted $S^{pred}$, the estimated RFS can be calculated with:

$$RFS^{pred} = \sum_{k=1}^{N} \left(\prod_{i=1}^{k} S_i^{pred}\right) \times T_k, \quad (2)$$

where $T \in \mathbb{R}^N$ is the durations of $N$ time intervals. The $RFS^{pred}$ was regarded as the survival score and was used to calculate the C-index during evaluation.

For the segmentation loss terms $\mathcal{L}_{PT}$ and $\mathcal{L}_{MLN}$, we sum the Dice loss and Focal loss:

$$\mathcal{L}_{PT/MLM} = \mathcal{L}_{Dice} + \mathcal{L}_{Focal}, \quad (3)$$

$$\mathcal{L}_{Dice} = \frac{2\sum_i^N p_i g_i}{\sum_i^N p_i^2 + \sum_i^N g_i^2}, \quad (4)$$

$$\mathcal{L}_{Focal} = -\sum_i^N \alpha g_i (1 - p_i)^\gamma \log(p_i) - (1 - g_i) p_i{}^\gamma \log(1 - p_i), \quad (5)$$

where $N$ is the sample size, $p$ is the output of the segmentation head, $p$ is the ground-truth segmentation labels. In the $\mathcal{L}_{Focal}$, $\alpha$ is a parameter for the trade-off between precision and recall (set to 0.25 as default), and $\gamma$ is a focusing parameter (set to 2 as default).

## B. Radiomics Feature Extraction

For radiomics enhancement, radiomics features are extracted from the XSurv-predicted PT/MLN regions of PET/CT images. The predicted PT and MLN masks are merged into a single mask (1 for PT/MLN and otherwise 0) for feature extraction. The extracted features include the features from First Order Statistics (FOS), Neighboring Grey Tone Difference Matrix (NGTDM), Grey-Level Run Length Matrix (GLRLM), Grey-Level Size Zone Matrix (GLSZM), Grey-Level Cooccurrence Matrix (GLCM), and 3D shape-based features. In addition to the original PET/CT, eight wavelet decompositions of PET/CT are also used, resulting in a total of 1689 radiomics features. All radiomics features are standardized using Z-score normalization.

## C. Visualization of Attention Maps

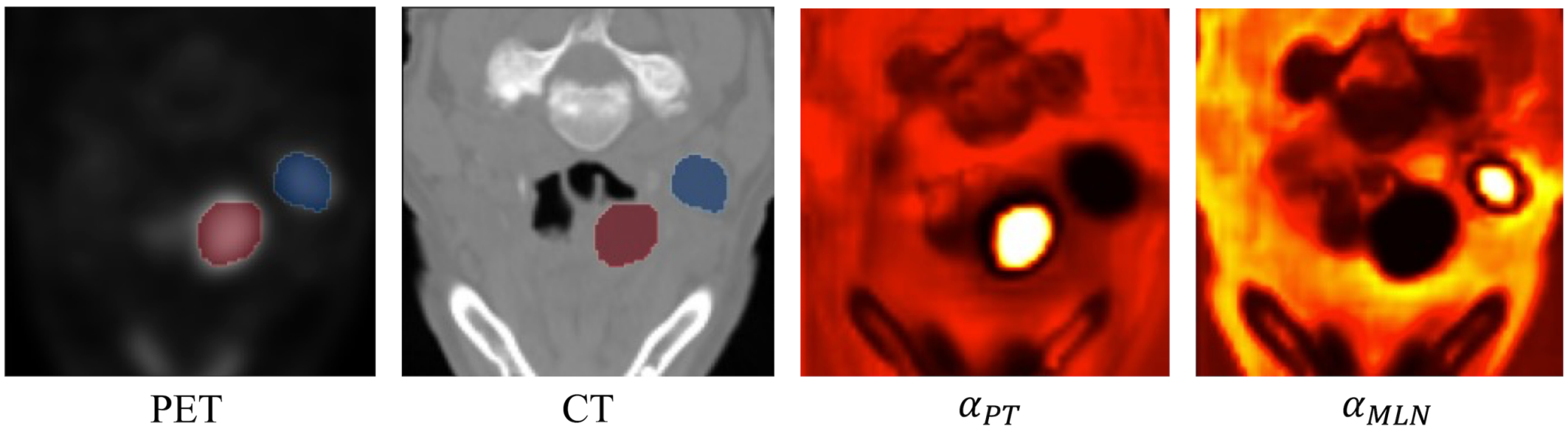

**Fig. S1.** Visualization of the attention maps $\alpha_{PT}$ and $\alpha_{MLN}$ produced in the diverging decoder. PET-CT images are presented with PT (red) and MLN (blue) ground-truth annotations.

## D. Architecture Details

**Table S1.** Architecture details of the XSurv used in the experiments.

| $C_e$ | $C_d$ | $n_{conv}$ | $n_{trans}$ | Head numbers | Window size |
|---|---|---|---|---|---|
| [8, 16, 32, 64, 128] | [128, 64, 32, 16, 8] | [2, 3, 3, 4, 4] | [0, 2, 2, 2, 2] | [0, 2, 4, 8, 16] | [5, 5, 5] |

$C_e$: the channel numbers/embedding dimensions used in the encoder. $C_d$: the channel numbers used in the decoder. $n_{conv}$ : The number of convolution operations. $n_{trans}$: The number of Swin Transformer blocks.

## E. Selection of Architecture Parameters

**Table S2.** Cross-validation results of the XSurv with different architecture parameters settings.

| $N_{conv}$ | $N_{self}$ | $N_{cross}$ | Survival prediction (C-index) | PT segmentation (DSC) | MLN segmentation (DSC) |
|---|---|---|---|---|---|
| 5 | 0 | 0 | 0.720 | 0.795 | 0.739 |
| 2 | 3 | 0 | 0.722 | 0.796 | 0.743 |
| 1 | 4 | 0 | 0.728 | 0.800 | 0.745 |
| 2 | 0 | 3 | 0.730 | 0.798 | 0.748 |
| 1 | 0 | 4 | 0.711 | 0.794 | 0.735 |
| 1 | 1 | 3 | **0.737** | **0.802** | **0.753** |

**Bold**: the best result in each column is in bold.

## F. Clinical Characteristics

**Table S3.** Clinical characteristics of patients in the training and testing sets.

| Characteristics | Training set | Testing set |
|---|---|---|
| Number of Patients | 386 | 102 |
| RFS, Number (%) | | |
| Uncensored | 81 (21.0) | 15 (14.7) |
| Censored | 305 (79.0) | 87 (85.3) |
| Age (year), median (range) | 60 (32-85) | 64 (44-90) |
| Weight (kg), median (range) | 81.5 (41-160) | 74 (34-120) |
| Gender, Number (%) | | |
| Male | 322 (83.4) | 80 (78.4) |
| Female | 64 (16.6) | 22 (21.6) |
| Alcohol consumption, Number (%) | | |
| Yes | 95 (24.6) | 0 (0.0) |
| No | 59 (15.3) | 0 (0.0) |
| Unknown | 232 (60.1) | 102 (100.0) |
| Tobacco consumption, Number (%) | | |
| Yes | 85 (22.0) | 0 (0.0) |
| No | 105 (27.2) | 0 (0.0) |
| Unknown | 196 (50.8) | 102 (100.0) |
| HPV status, Number (%) | | |
| Positive | 252 (65.3) | 22 (21.6) |
| Negative | 41 (10.6) | 2 (2.0) |
| Unknown | 93 (24.1) | 78 (76.4) |
| Performance status, Number (%) | | |
| 0 | 86 (22.3) | 0 (0.0) |
| 1 | 114 (29.5) | 0 (0.0) |
| 2 | 11 (2.8) | 0 (0.0) |
| 3 | 3 (0.8) | 0 (0.0) |
| 4 | 1 (0.3) | 0 (0.0) |
| Unknown | 171 (44.3) | 102 (100.0) |
| Surgery, Number (%) | | |
| Yes | 50 (13.0) | 0 (0.0) |
| No | 202 (52.3) | 46 (45.1) |
| Unknown | 134 (34.7) | 56 (54.9) |
| Chemotherapy, Number (%) | | |
| Yes | 324 (83.9) | 98 (96.1) |
| No | 62 (16.1) | 4 (3.9) |